\documentclass[reprint,superscriptaddress,amsmath,amssymb,aps,dvipsnames,usenames]{revtex4-2}

\usepackage[dvipsnames]{xcolor}
\usepackage{graphicx}
\usepackage{dcolumn}
\usepackage{bm}
\usepackage{amsmath}
\usepackage{notes2bib}
\newcommand{\highlight}[1]{%
  \colorbox{gray!20}{$\displaystyle#1$}}

\begin{document}

\title{Unifying frequency combs in active and passive cavities:\\
Temporal solitons in externally-driven ring lasers}

\author{L. Columbo$^\dagger$}
\affiliation{Dipartimento di Elettronica e Telecomunicazioni, Politecnico di Torino,
Torino, Italy}
\affiliation{CNR-Istituto di Fotonica e Nanotecnologie, Bari, Italy}

\author{M. Piccardo$^\dagger$}
\email{piccardo@g.harvard.edu}
\affiliation{Center for Nano Science and Technology, Fondazione Istituto Italiano di Tecnologia,
Milano, Italy}
\affiliation{Harvard John A. Paulson School of Engineering and Applied Sciences, Harvard University, Cambridge, MA, USA}

\author{F. Prati}
\affiliation{Dipartimento di Scienza e Alta Tecnologia, Universit\`a dell'Insubria,
Como, Italy}

\author{L.A. Lugiato}
\affiliation{Dipartimento di Scienza e Alta Tecnologia, Universit\`a dell'Insubria,
Como, Italy}

\author{M. Brambilla}
\affiliation{Dipartimento di Fisica Interateneo and CNR-IFN, Universit\`a e Politecnico di Bari, 
Bari, Italy}

\author{A. Gatti}
\affiliation{Istituto di Fotonica e Nanotecnologie IFN-CNR,
Milano, Italy}
\affiliation{Dipartimento di Scienza e Alta Tecnologia, Universit\`a dell'Insubria,
Como, Italy}

\author{C. Silvestri}
\affiliation{Dipartimento di Elettronica e Telecomunicazioni, Politecnico di Torino,
Torino, Italy}

\author{M. Gioannini}
\affiliation{Dipartimento di Elettronica e Telecomunicazioni, Politecnico di Torino,
Torino, Italy}

\author{N. Opa\v{c}ak}
\affiliation{Institute of Solid State Electronics, TU Wien, Vienna, Austria}

\author{B. Schwarz}
\affiliation{Institute of Solid State Electronics, TU Wien, Vienna, Austria}

\author{F. Capasso}
\affiliation{Harvard John A. Paulson School of Engineering and Applied Sciences, Harvard University, Cambridge, MA, USA}

\collaboration{$^\dagger$These authors contributed equally to this work.}

\begin{abstract}
Frequency combs have become a prominent research area in optics. Of particular interest as integrated comb technology are chip-scale sources, such as semiconductor lasers and microresonators, which consist of resonators embedding a nonlinear medium either with or without population inversion. Such active and passive cavities were so far treated distinctly. Here we propose a formal unification by introducing a general equation that describes both types of cavities. The equation also captures the physics of a hybrid device---a semiconductor ring laser with an external optical drive---in which we show the existence of temporal solitons, previously identified only in microresonators, thanks to symmetry breaking and self-localization phenomena typical of spatially-extended dissipative systems.
\end{abstract}
\date{\today}

\maketitle

\begin{figure*}[th!]
    \centering
    \includegraphics[width=0.8\textwidth]{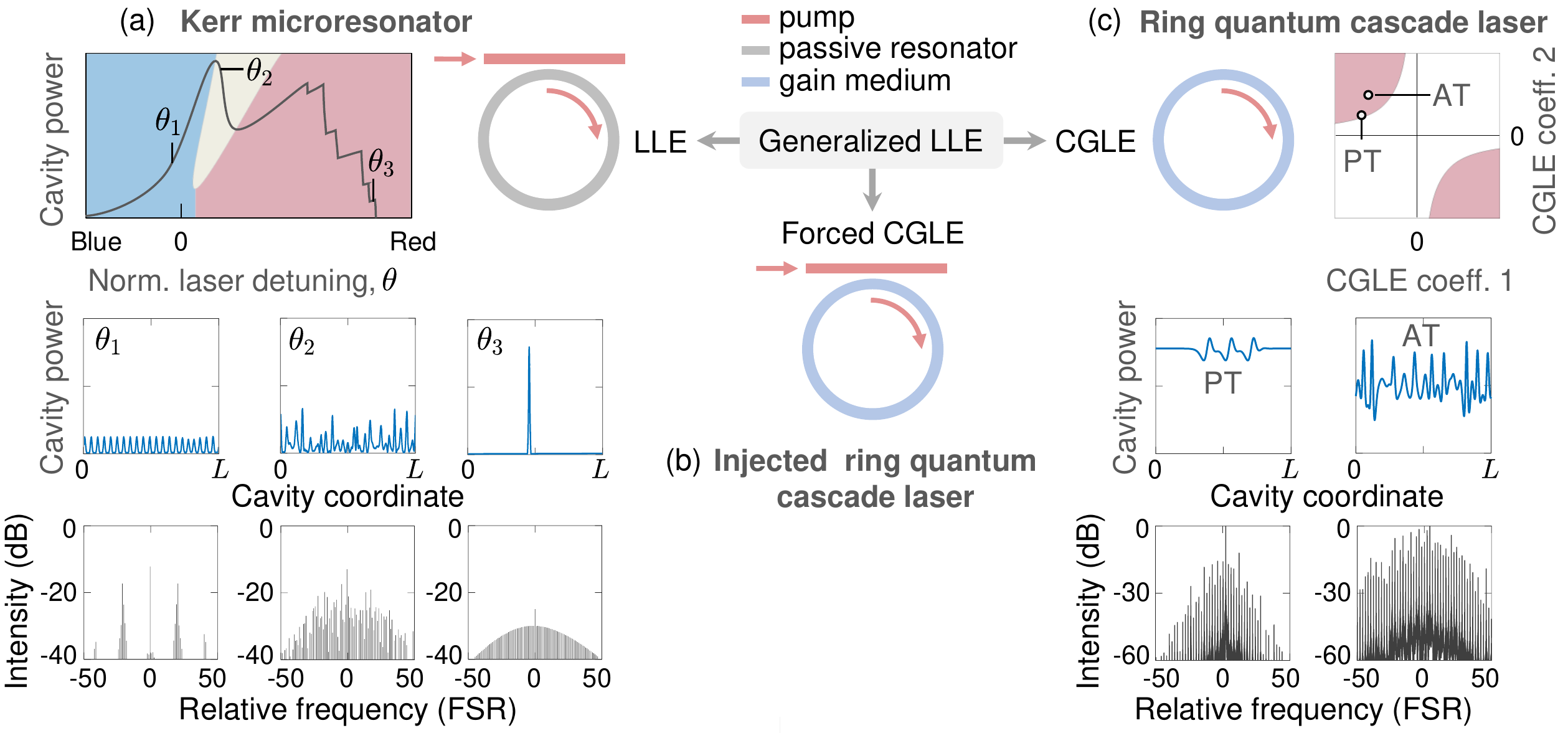}
    \caption{\textbf{Active and passive frequency comb sources: unification.} (a) Schematic of a passive microresonator described by the LLE. The intracavity power depends on the pump laser detuning $\theta$ resulting in regions with different patterns: Turing rolls (blue), breather solitons (yellow), and stable temporal solitons (red). An example of different intensity distributions along the cavity and corresponding comb spectrum is given in each case. $L$ denotes the cavity length. Images adapted from \cite{Herr2014}. (b) Schematic of an intermediate system between the active and passive case consisting of a ring quantum cascade laser (QCL) injected by an external optical signal, which is described by a forced complex Ginzburg-Landau equation (CGLE). (c) Schematic of a unidirectional ring QCL described by the CGLE, which depends on two coefficients, related to the linewidth enhancement factor and group velocity dispersion of the laser. The ring laser can undergo a multimode transition in the red regions of the parameter space. The resulting dynamic behavior can be characterized by phase (PT) or amplitude (AT) turbulence.}
    \label{fig_extendedLLE}
\end{figure*}

\textit{Introduction}---The discovery of optical frequency combs \cite{Udem2002,Jones2000} (OFCs) in high-Q ring microresonators filled with a Kerr medium, such as SiO$_2$, Si$_3$N$_4$ and diamond \cite{Kippenbergeaan8083}, and driven by an external laser beam activated worldwide attention on Kerr frequency combs (KFCs), because this avenue offers substantial potential for miniaturization and chip-scale photonic integration \cite{DelHaye2007,Shen2020}. This technology has been applied to numerous areas, including coherent telecommunications and laser ranging \cite{Kippenberg2011,Chembo2016}. It was recognized later \cite{Chembo2013,Herr2014} that the physics of KFCs corresponds very accurately to the model formulated in 1987 by the Lugiato-Lefever equation (LLE) \cite{Lugiato1987}. This is a one-dimensional nonlinear Schrödinger equation in the presence of an external driving, linear damping and detuning. The spontaneous formation of spatial patterns travelling along the cavity, described in the LLE, is the spatio-temporal equivalent of the frequency combs and governs their features \cite{Lugiato2018}. By varying the frequency detuning $\theta$ of the pump laser injecting the microresonator a variety of spatial patterns can form (Fig. \ref{fig_extendedLLE}(a)), such as Turing rolls, breather solitons and stable temporal solitons. A common feature of these spectra is that their envelope is bell-shaped and can be approximated by a hyperbolic-secant function (sech) \cite{Coen2013a,Kippenbergeaan8083}.



Recently, frequency comb spectra with sech-type envelope were also observed in ring quantum cascade lasers (QCLs, Fig. \ref{fig_extendedLLE}(c)) \cite{Piccardo2020,Meng:20}. These are unipolar semiconductor lasers, first realized in 1994 \cite{Faist1994,faistbook2013}, emitting in the mid-infrared and terahertz regions of the electromagnetic spectrum. QCLs have attracted much attention, especially in mid-infrared spectroscopy and sensing \cite{Hugi2012,Villares2014}, thanks to their tunability via band-gap engineering and unique physical properties \cite{Choi2008PRL}, such as ultrafast gain dynamics and strong resonant third-order nonlinearity. The study of ring QCLs operating in a unidirectional regime revealed a number of similarities with KFCs. It was found that the multimode laser instability is produced by the interplay of dispersive and nonlinear effects, as in the modulational instability (MI) of passive microresonators, and the number of localized structures appearing in the spatial patterns varies stochastically with the initial conditions \cite{Piccardo2020}---a phenomenon known as multistability, also occurring in KFCs \cite{Kippenbergeaan8083}.


The similarity between the behaviours of these two disparate sources 
can be traced back to a precise  formal analogy between their model equations, outlined by the scheme of  Fig. \ref{fig_extendedLLE}. Indeed, it has been shown \cite{Piccardo2020} that, under conditions of fast material dynamics and  near-threshold operation, the dynamics of the ring QCL is well described by a complex Ginzburg-Landau equation (CGLE), where the two coefficients of the equation (Fig. \ref{fig_extendedLLE}(c)) are determined by the linewidth enhancement factor \cite{Chow} (LEF or $\alpha$-factor) of the laser and by its group velocity dispersion (GVD) \cite{Opacak2019}.
A CGLE was also formulated in \cite{Lugiato1988}, and analyzed in \cite{Wang1993}, in order to describe the 2D patterns arising in the transverse plane of the resonator due to the interplay of nonlinearity and diffraction. Such an equation can be regarded as the active counterpart of the LLE, which describes patterns in a driven passive resonator. Since the LLE, restricted to 1D, also describes pulses travelling along the longitudinal axis of the resonator \cite{Haelterman1992,Castelli2017} it is natural to think of a similar equation for the active case.

These considerations led us to unify the two contexts of frequency combs in passive and active systems by formulating a generalized LLE, defined as the simplest equation that includes the passive and the active LLE as special cases. In turn, this step quite naturally leads to envisage a novel configuration, namely a ring QCL with injected signal (Fig. \ref{fig_extendedLLE}(b)), which is studied in this work.

\textit{Generalized Longitudinal LLE}---Let us consider the following equation that describes the spatiotemporal evolution of the envelope $E$ of the electric field in an optical cavity
\begin{eqnarray}\label{eq:gen}
    \tau_p\:\partial_t E&=& \mathop{\highlight{E_I}}^{\mathrm{driving}} + \mathop{\highlight{\left(-1-i\theta_0\right)E}}^{\mathrm{damping-detuning}}
    + \mathop{\highlight{\left(d_R+id_I\right)\partial_z^2E}}^{\mathrm{diffusion-dispersion}} \nonumber \\
    &&+\mathop{\highlight{\mu\left(1-i\Delta\right)\left(1-|E|^2\right)E}}_{\mathrm{gain/loss,~nonlinearity}}\:,
\end{eqnarray}
%
where $t$ and $z$ are the temporal and spatial coordinate along the cavity axis, in a reference  frame moving at the light velocity in the cavity, and $\tau_p$ is the damping time of the cavity field. Electric fields are scaled \cite{Columbo2018,NOS} to present the equation in its simplest form (Supplementary Material). $E_I$ is the amplitude of a coherent field injected in the cavity, which may or may not be present. In the second term, the $-1$ accounts for cavity losses and $\theta_0$ is a detuning parameter. 
In the third term, the differential operator applied to cavity modes $\propto \mathrm{e}^{ik_n z} E_n$ provides an algebraic term $-id_Ik_n ^2E_n-d_R k_n^2E_n$ whose imaginary part is associated with frequency dispersion while the real part (for $d_R>0$) is a diffusion term that acts as a cut-off on the frequency spectrum. This term is connected to the reaction-diffusion mechanism responsible for pattern formation as described in Turing's theory of morphogenesis \cite{Turing1952}, where in our case the reaction is produced by all the other linear and nonlinear terms appearing in Eq. (\ref{eq:gen}).  Both terms  arise from an adiabatic elimination of the material variables (Supplementary Material, which includes Ref. \cite{Prati2020}) that takes into account the fast but not instantaneous response of the medium.  In addition, $d_I$ may contain the contribution of the GVD of a host medium. The fourth term describes the linear and nonlinear interaction of the electric field with the medium, as obtained by an adiabatic elimination of the material variables under the approximation $|E|^2\ll 1$. In this term, $\mu$ is the unsaturated gain ($\mu>0$) or absorption ($\mu<0$) parameter. The coefficient of the nonlinearity $\Delta$ depends on the system under consideration, being e.g. the atomic detuning  for  two-level media, or the LEF in the case of semiconductor lasers. In the latter case, it can also contain a contribution  from  the Kerr nonlinearity of the host medium.
In the following we will concentrate on the different limits of Eq. (\ref{eq:gen}). Let us consider the passive and active case in order.

\textit{Passive case}--- Here we assume that the medium is a weak absorber ($\mu<0$, $|\mu|\ll1$) with strong, negative atomic detuning ($\Delta<0$, $|\Delta|\gg1$) and large resonance curve bandwidth ($d_I\gg d_R$), so that the approximations $\mu(1-i\Delta)\approx -i\mu\Delta$ and $d_R+id_I\approx id_I$ hold true and we obtain the LLE
\begin{equation}
    \partial_\tau F=F_I-\left[1+i\left(\theta-|F|^2\right)\right]F
    +i\partial^2_\eta F\,,
\end{equation}
with $\theta=\theta_0+\mu\Delta$, $F=\sqrt{\mu\Delta}\,E$, $F_I=\sqrt{\mu\Delta}\,E_I$, $\tau=t/\tau_p$, and $\eta=z/\sqrt{d_I}$, where we assume that the dispersion is anomalous so that $d_I >0$. In this case $\theta_0$ must be taken as the cavity detuning $\theta_c=(\omega_c-\omega_0)\tau_p$, $\omega_c$ being the empty  cavity frequency closest to the frequency $\omega_0$ of the incident field.


\textit{Active case}--- Here we assume that the medium is active and close to the lasing threshold ($\mu=1+r$, $|r|\ll1$, above threshold for $r>0$ and below for $r<0$) and that $|E|^2$ has the same order of magnitude as $|r|$ so that the approximation $\mu(1-|E|^2)\approx\mu-|E|^2$ is justified and we obtain an equation formally equivalent to a forced CGLE for
\begin{equation}\label{eq:FCGLE}
    \partial_\tau F =F_I+\gamma\left(1-i\theta\right)F
    -\left(1-i\Delta\right)|F|^2F  
    +\left(1+iG\right)\partial^2_\eta F\,,
\end{equation}
with $\gamma=r/|r|$, $\theta=(\theta_0+\mu\Delta)/r$, $G=d_I/d_R$, $F=E/\sqrt{|r|}$, $F_I=E_I/|r|^{3/2}$, $\tau=|r|t/\tau_p$, and $\eta=z\sqrt{|r|/d_R}$. 
The relevant parameter 
is the detuning of the frequency $\omega_0$ of the injected signal with respect to the frequency $\omega_L$ of the solitary laser. We show (Supplementary Material) that this is  given by  
$(\omega_L-\omega_0)\tau_p=r(\theta-\Delta)$.

In the Supplemental Material we show that an equation identical to Eq. (\ref{eq:FCGLE}) can be derived from a full laser model for a QCL with coherent injection in the limit of ultrafast carriers and in proximity of the lasing threshold. 
In this case we have $\gamma=1$ and 
\begin{equation}\label{eq:abz}
\Delta=\alpha+\beta\,,\;
G=\alpha+\zeta\,,\quad \zeta=-(1+\alpha^2)\frac{\tilde{c}\tau_p}{2\tau_d^2}k''\,,
\end{equation}
where $\alpha$ is the linewidth enhancement factor (LEF) \cite{Henry1982}, $\beta$ and $k''$ are the Kerr and the the GVD coefficient, respectively, of the host medium, while $\tilde{c}$ and $\tau_d$ are the speed of light and the polarization dephasing time in the QCL with group index $n$.
In this case $\theta_0=\theta_c-\mu\beta$ and therefore
$\theta=(\theta_c+\mu\alpha)/r$.

Above threshold and without an injected field Eq. (\ref{eq:FCGLE}) with $\theta=0$, $\Delta=c_\mathrm{NL}$ and $G=-c_\mathrm{D}$ coincides with the CGLE in \cite{Piccardo2020}  \bibnote{In \cite{Lugiato1987} and \cite{Lugiato1988} and in this paper the field envelope $E$ multiplies the factor $\mathrm{e}^{-i\omega_0t}$, whereas in \cite{Piccardo2020} it multiplies the factor $\mathrm{e}^{i\omega_0t}$, so that the $E$ which appears in \cite{Piccardo2020} is the complex conjugate of the $E$ which appears in this paper.}.

The dynamics of the electric field in the transverse plane of a similar system, i.e. a class-A laser with injected signal, was studied in \cite{Gibson2016} using an equation like Eq. (\ref{eq:FCGLE}), with the second order derivative along $\eta$ replaced by the transverse Laplacian with a purely imaginary coefficient and $\Delta=0$. Equation (\ref{eq:FCGLE}) with $\Delta=G=\alpha$ was already successful in describing the formation of phase solitons occurring in a driven bipolar semiconductor ring laser with a meter-size extended cavity \cite{PRLgustave}. Besides the ring geometry considered here, a connection between QCLs and the LLE was also established recently in the case of Fabry-Perot devices \cite{Burghoff2020Optica}.

Temporal cavity solitons without background in a passive microcavity coupled with an amplifying fiber loop were demonstrated in \cite{Bao2019}. This system cannot be described by the generalized LLE because it requires two coupled equations.

\begin{figure*}[t!]
    \centering
    \includegraphics[width=1\textwidth]{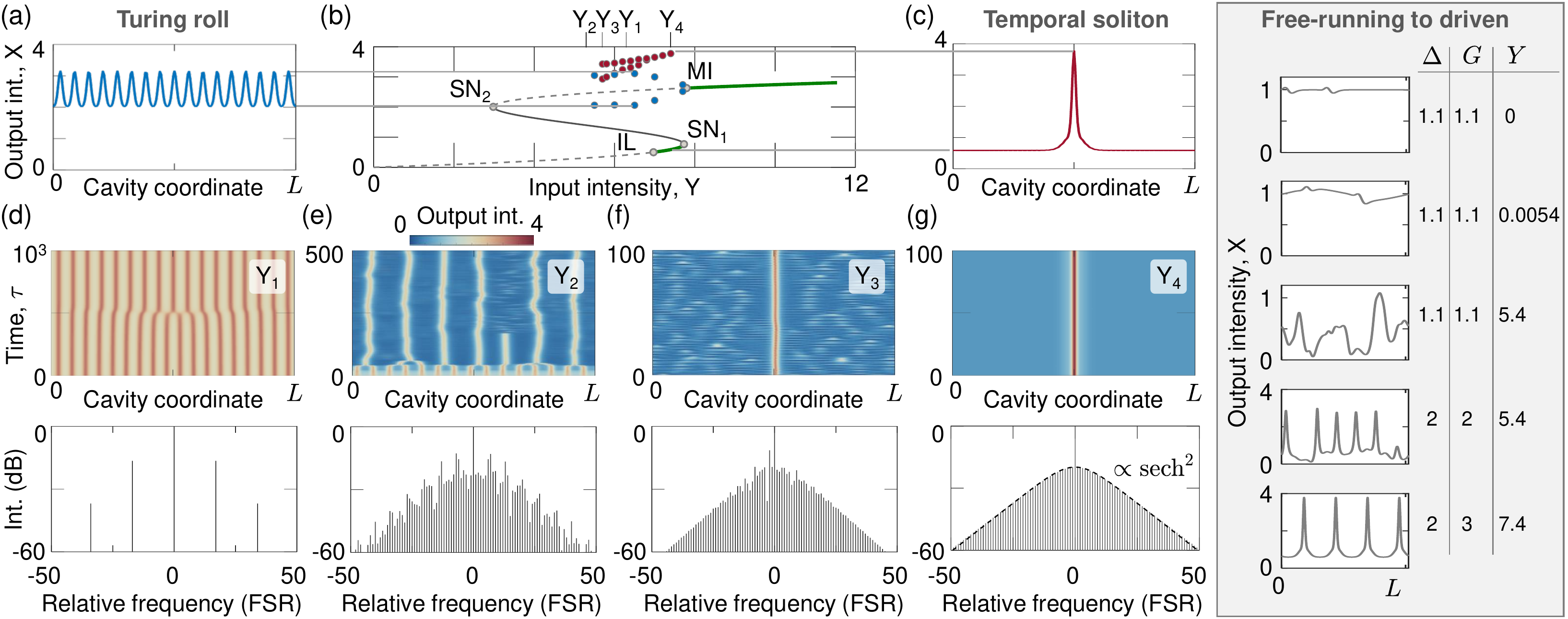}
    \caption{\textbf{Spatiotemporal dynamics of the injected ring quantum cascade laser.} (a) 1D Turing rolls exhibiting periodic oscillations between two  intensities,  corresponding to a pair of blue dots in (b). (b) S-shaped curve of output intensity $X=|F|^2$ vs. input intensity $Y=F_I^2$. Injection locking threshold (IL), saddle nodes (SN$_1$ and SN$_2$), modulation instability threshold (MI). Different segments of the curve can be stable (green line), unstable (dashed line) or not accessible (grey line). Blue and red dots correspond to Turing patterns and cavity solitons (CSs). (c) A CS with a pedestal and peak intensity corresponding to a point on the stable lower branch of the S-shaped curve and a red dot of (b), respectively. (d)--(g) Spatio-temporal plots and corresponding comb spectra of: (d) Turing roll; 
    (e) transition from Turing roll to non-stationary CSs;
    (f) CS on an unstable background; (g) stable CS.
    The input intensity is varied from $Y_1$ to $Y_4$ as marked in (b), i.e. first decreasing and then increasing again the intensity. Frequencies are relative to the central mode. Time is scaled to $\tau_p/|r|$, so that a smaller distance from threshold corresponds to a slower dynamics. The box shows from top to bottom localized structures emerging from phase instability in the free-running QCL \cite{Piccardo2020}, phase turbulence, amplitude turbulence, filamentation, CSs (see the full spatiotemporal plots in Supplementary Material), when the laser is driven along an arbitrary trajectory across the parameter space.}
    \label{fig_spontaneous_solitons}
\end{figure*}

\textit{The injected ring QCL}---When an external coherent field is injected into the QCL (Fig. \ref{fig_extendedLLE}(b)), the generalized LLE describing this configuration encompasses two more control parameters: the external field intensity and frequency. Moreover, the injection of an external field allows the stationary and homogeneous solution to assume an S-shape, as shown by Fig. \ref{fig_spontaneous_solitons}(b)---a phenomenon known as optical bistability, occurring also in passive Kerr microresonators.
In the active case above threshold only a segment of the lower branch of this curve is stable, and this occurs between the injection locking (IL) point, where the lasing frequency is locked to the injected field, and the turning point SN$_1$ (green segment in Fig. \ref{fig_spontaneous_solitons}(b)). At the same time, part of the upper branch of the curve is affected by a MI---a spontaneous symmetry breaking mechanism producing intensity patterns characterized by a high degree of spatial correlation (Fig. \ref{fig_spontaneous_solitons}(d)), or else spatiotemporal turbulence \cite{Aranson2002}. As in passive microresonators, the S-shape of the stationary curve creates conditions favorable for the generation of temporal solitons, referred to also as cavity solitons (CSs), i.e. dissipative localized structures formed inside an optical resonator. These conditions correspond to having an interval of input intensities in which the upper branch of the S-curve is modulationally unstable and coexists with a stable homogeneous state in the lower branch.

%

In this situation the system might form a localized pattern emerging from the MI on a uniform stable background, and eventually give origin to a CS \cite{NOS} (see Fig. \ref{fig_spontaneous_solitons}(c,g)). These considerations guided our search for CSs (and the associated OFCs) in QCLs with injected signal and in particular the choice of an experimentally reasonable parameter set. Fig. \ref{fig_spontaneous_solitons} shows the results of numerical simulations of Eq. \eqref{eq:FCGLE}, performed with the following parameters:
$\gamma=1$, $\theta=4.7$, $\alpha=2$, $\beta=0$, $\zeta=1$ ($\Delta=2$, $G=3$). By assuming $n=3.3$, $\tau_p = 50$ ps and $\tau_d = 60$ fs \cite{Piccardo2020}, our parameters correspond to a slight red-detuning of the pump field, e.g. 0.9 GHz for a laser 10$\%$ above threshold, and the GVD calculated from Eq. (\ref{eq:abz}) turns out around $-300$ fs$^2/$mm, a realistic value for QCLs \cite{Kazakov2017}.
The S-shaped curve calculated for the selected parameter set is shown in Fig. \ref{fig_spontaneous_solitons}(b).
Note that, while the plotted input and output intensity are comparable, the corresponding physical quantities are scaled to $r^{3}$ and $r$, respectively, so that the physical injected intensity is much smaller than the output one.

By varying the input intensity $Y=F^2_I$
\bibnote{Taking $F_I$ real merely sets the reference phase for the optical field}, we observe the following scenario emerging from numerical integration \bibnote{We specify that a stochastic noise term mimicking spontaneous emission in the system was included in all the simulations.} of Eq. (\ref{eq:FCGLE}). 
Starting on the high-intensity spatially-uniform solution, stable to the right of the MI point, and progressively decreasing $Y$, a globally modulated pattern bifurcates from MI at $Y_\mathrm{MI}=7.8$ (Fig. \ref{fig_spontaneous_solitons}(a) shows the pattern for $Y_1=6.3$). The bifurcation is by definition supercritical as the branch that bifurcates remains stable, down to about $Y=5.4$. The modulated pattern corresponds to a 1D Turing roll \cite{Turing1952} and its branch is indicated by the blue dots in Fig. \ref{fig_spontaneous_solitons}(b), which mark the maximum and minimum intensity. 
The period of the spatial modulation of the roll pattern depends on $Y$. This feature makes the injected ring QCL particularly appealing, because the comb spacing can be tuned by simply changing the intensity of the injected signal, rather than by widely tuning its frequency as it was done for the control of the harmonic state in Fabry-Perot QCLs \cite{Piccardowidelytunableharmonic}. Figure \ref{fig_spontaneous_solitons}(d), in fact, shows a simulation where a period-$18$ roll pattern (stable only at $Y=6.4$), taken as an initial condition, spontaneously evolves to a stable period-$17$ roll when $Y=Y_1=6.3$.


Below $Y=5.4$ the rolls become unstable and the system undergoes a spontaneous collapse of the roll pattern. It evolves into a number of non-stationary CSs sitting on a turbulent background, which corresponds to the unstable lower branch of the steady-state curve (Fig. \ref{fig_spontaneous_solitons}(e), $Y_2=5.3$).
A further decrease of $Y$ brings the system in a turbulent regime where any ordered structure disappears. Conversely, by starting from the non-stationary CS and increasing the input intensity in the interval $5.7\le Y<7$ a single CS with turbulent background survives (Fig. \ref{fig_spontaneous_solitons}(f), $Y_3=5.8$). The background fluctuations cause a jitter in the soliton shape and intensity maximum. The range of fluctuations of the CS peak are traced by the pairs of red dots in Fig. \ref{fig_spontaneous_solitons}(b). Finally, by following upwards the soliton branch in the interval $7\le Y\le7.4$ the pedestal of the single CS becomes stable corresponding to the lower uniform and stable branch of the steady-state curve, as expected, since $Y_\mathrm{IL}=6.97$ (Fig. \ref{fig_spontaneous_solitons}(c,g), $Y_4=7.4$). 
The CS spatial shape and corresponding spectra, well approximated by a sech$^2$ envelope (Fig. \ref{fig_spontaneous_solitons}(f,g)), do not change in time. We note that CSs are not only predicted by our reduced model, but are also observed in our full dynamical model of the QCL (see Eqs. (S1)-(S3) of the Supplementary Material, which includes Refs. \cite{Prati2007,Li2015}). Moreover, we show that CSs emerge also when transitioning from the free-running \cite{Piccardo2020} to driven case, when an appropriate path is swept in the multidimensional parameter space (Fig. \ref{fig_spontaneous_solitons}, box).

The most appealing  features of CSs, from the applicative viewpoint, are multistability, independence and plasticity. The CSs reported here are: a) stable versus the considerable fluctuations such as those shown in the background of Fig. \ref{fig_spontaneous_solitons}(f)); b) intrinsically multistable so that the injection of short pulses allows to excite multiple CSs (see Supplementary Material, which includes Ref. \cite{Brambilla1996}, for related simulations).

\textit{Conclusions}---The generalized LLE introduced in this work makes it possible to connect for the first time from a formal viewpoint Kerr microresonators and QCLs. The injected ring QCL is a direct result of this unification opening a pathway for the realization of new spatiotemporal patterns in QCLs such as Turing rolls and CSs, previously restrained to Kerr combs. The CS emerging from a CW input field demonstrate the possibility of generating high-contrast short pulses in this device. Although this result comes as a surprise, as it has long been thought that the ultrafast dynamics of QCLs should strongly suppress amplitude modulation \cite{PiccardoVariational} in absence of a radiofrequency modulation of the gain \cite{HillbrandPulses2019}, here we have shown that a short pulse regime is possible, thanks to soliton formation triggered by a compensation between dispersion and nonlinear self-phase modulation associated with finite LEF in a resonator with gain driven by an external CW optical signal. In \cite{PRLgustave} this requirement was fulfilled using a long external cavity ($\approx 1$ m) and a standard semiconductor bipolar laser with nanosecond gain recovery time. In this work we showed that unipolar lasers (QCLs), having carrier dynamics 3 orders of magnitude faster, allow to downsize the cavity length to the millimeter range, with significant impact on chip-scale frequency comb applications.


%

\end{document}


\title{Supplementary Material to:\\
Unifying frequency combs in active and passive cavities:\\
Temporal solitons in externally-driven ring lasers}

\author{L. Columbo$^\dagger$}
\affiliation{Dipartimento di Elettronica e Telecomunicazioni, Politecnico di Torino,
Torino, Italy}
\affiliation{CNR-Istituto di Fotonica e Nanotecnologie, Bari, Italy}

\author{M. Piccardo$^\dagger$}
\email{piccardo@g.harvard.edu}
\affiliation{Center for Nano Science and Technology, Fondazione Istituto Italiano di Tecnologia,
Milano, Italy}
\affiliation{Harvard John A. Paulson School of Engineering and Applied Sciences, Harvard University, Cambridge, MA, USA}

\author{F. Prati}
\affiliation{Dipartimento di Scienza e Alta Tecnologia, Universit\`a dell'Insubria,
Como, Italy}

\author{L.A. Lugiato}
\affiliation{Dipartimento di Scienza e Alta Tecnologia, Universit\`a dell'Insubria,
Como, Italy}

\author{M. Brambilla}
\affiliation{Dipartimento di Fisica Interateneo and CNR-IFN, Universit\`a e Politecnico di Bari, 
Bari, Italy}

\author{A. Gatti}
\affiliation{Istituto di Fotonica e Nanotecnologie IFN-CNR,
Milano, Italy}
\affiliation{Dipartimento di Scienza e Alta Tecnologia, Universit\`a dell'Insubria,
Como, Italy}

\author{C. Silvestri}
\affiliation{Dipartimento di Elettronica e Telecomunicazioni, Politecnico di Torino,
Torino, Italy}

\author{M. Gioannini}
\affiliation{Dipartimento di Elettronica e Telecomunicazioni, Politecnico di Torino,
Torino, Italy}

\author{N. Opa\v{c}ak}
\affiliation{Institute of Solid State Electronics, TU Wien, Vienna, Austria}

\author{B. Schwarz}
\affiliation{Institute of Solid State Electronics, TU Wien, Vienna, Austria}

\author{F. Capasso}
\affiliation{Harvard John A. Paulson School of Engineering and Applied Sciences, Harvard University, Cambridge, MA, USA}

\collaboration{$^\dagger$These authors contributed equally to this work.}

\maketitle

\section{\normalsize{E\lowercase{ffective} S\lowercase{emiconductor} M\lowercase{axwell}-B\lowercase{loch} E\lowercase{quations for a multimode} QCL \lowercase{with injected signal}}}

Recently, a set of effective semiconductor Maxwell-Bloch equations (ESMBEs) originally introduced in \cite{Prati2007} was successfully applied  \cite{Columbo2018} to describe the coherent multimode dynamics of a quantum cascade laser (QCL). Contrary to the standard Maxwell-Bloch approach \cite{faistbook2013}, the ESMBEs allow to self-consistently and properly reproduce typical experimental observations of self-starting optical frequency combs (OFCs) in the system together with the alternation among regular and irregular dynamics \cite{Li2015} by sweeping the bias current. The model encompasses a nonlinear optical susceptibility that accounts for peculiar features of radiation-matter interaction in semiconductor lasers such as asymmetric gain/dispersion curves and phase-amplitude coupling due to the linewidth enhancement factor (LEF) \cite{Chow}. Interestingly, in \cite{Columbo2018} the existence in a unidirectional ring configuration (where spatial hole burning due to carriers grating cannot occur) of a continuous wave instability close to the lasing threshold was demonstrated---an original result in perfect agreement with very recent experimental evidences \cite{Piccardo2020}.
In the following we extend the ESMBEs of \cite{Columbo2018} to include a detuned, coherent driving field; assuming fast medium variables and a laser close to threshold, we derive the generalized longitudinal LLE equation (Eq. (1) of the main text in the active case).

The set of ESMBEs in presence of an injected signal $E_{I}$, in the additional hypothesis of gain peak coincident with the reference frequency (frequency of the injected field) is
%
\begin{eqnarray}
\tilde{c} \frac{\partial E}{\partial z}+\frac{\partial E}{\partial t} &=&\frac{1}{\tau_{p}} \left[-(1+i \theta_c)E+E_{I}+P+ i\beta |E|^2E-i\tilde{c}\tau_p\frac{k''}{2}\frac{\partial^2 E}{\partial t^2}\right]\label{E4}\\
\frac{\partial P}{\partial t}&=&\frac{1}{\tau_{d}} \left(1-i \alpha \right)\left[(1-i\alpha)E D-P \right]   \label{P4a}\\
\frac{\partial D}{\partial t}&=&\frac{1}{\tau_{e}}\left [ \mu-D-\frac{1}{2}\left(E^{*}P+ E P^{*} \right)\right]\label{P4}
\end{eqnarray}
where $\tau_{p}$, $\tau_{d}$ and $\tau_{e}$ are the damping time of the cavity field, the dephasing time and the carriers lifetime respectively, and $\mu$ is the pump parameter scaled to its threshold value. 
The damping time of the cavity field $\tau_p$ is defined as $\tau_p=2L/(\tilde{c}T_\mathrm{eff})$
where the effective loss parameter
$T_\mathrm{eff}=T+\alpha_{L}L$ accounts for both localized intensity transmission losses ($T$), and distributed waveguide intensity absorption losses ($\alpha_{L}L$).
The variables $E$, $P$ and $D$ are scaled as in \cite{Columbo2018} while $E_{I}$ is introduced as in \cite{NOS}.

In Eq. (\ref{E4}) $\tilde{c}=c/n$ is the group velocity in the active medium, while $\theta_c=(\omega_c-\omega_0)\tau_p$ 
where $\omega_0$ is the reference frequency equal to the frequency of the injected field and $\omega_c$ is the closest cavity resonance. The $\alpha$ factor in Eq. (\ref{P4a}) is the LEF calculated at the gain peak.
Finally, $\beta$ and $k''$ are the Kerr nonlinearity parameter and the second-order dispersion coefficient respectively \cite{Opacak2019}.\\
Equations (\ref{E4})-(\ref{P4}) satisfy the periodic boundary condition
\begin{equation}
E(0,t)=E(L,t)\,. \label{BCnew}
\end{equation}
The formal solutions of Eqs. (\ref{P4a}) and (\ref{P4}) are
\begin{eqnarray}
P&=& (1-i\alpha)
\left(1+\frac{\tau_{d}}{1-i\alpha}
\frac{\partial}{\partial t}\right)^{-1}E D\,,\\
D&=&\mu -\frac{1}{2}\left(1+\tau_{e}
\frac{\partial }{\partial t}\right)^{-1}\left(E^*P+EP^*\right)\,.\label{eq:Dformal}
\end{eqnarray}
In the realistic hypothesis that the field evolves on the time scale of $\tau_{p}$ much bigger than those of the medium $\tau_{d}$ and $\tau_{e}$ (fast polarization and carrier dynamics), i.e. assuming 
$\frac{\partial }{\partial t}\ll \frac{1}{\tau_{d}}$ and $\frac{\partial }{\partial t}\ll \frac{1}{\tau_{e}}$, the time operators can be expanded in power series. 
For the polarization $P$ we must include terms up to second order in the expansion in order to keep information about the shape of the susceptibility, and we obtain
\begin{equation}
P\simeq(1-i\alpha)E D-\tau_{d}\frac{\partial (ED)}{\partial t}+\frac{\tau_{d}^{2}}{(1-i\alpha)}\frac{\partial^{2} (ED)}{\partial t^{2}} \label{P1}
\end{equation}
%
%
It turns out that the temporal operator in Eq. (\ref{eq:Dformal}) applies to a quantity of order $|E|^2$. 
If the laser is very close to threshold $|E|^2\ll1$, we can then keep only the zero order term of the operator and we simply get
\begin{equation} \label{Dnew}
D\simeq \mu(1-|E|^{2})\simeq \mu-|E|^{2}\,,
\end{equation}
where we have taken into account that $\mu \simeq 1$ (near threshold operation), and insert it in Eq. (\ref{P1}) to finally obtain
\begin{equation} \label{Pnew}
P\simeq(1-i\alpha)E  (\mu-|E|^{2}) 
-\tau_d\frac{\partial E}{\partial t}
+\frac{\tau_d^2}{(1-i\alpha)}
\frac{\partial^2 E}{\partial t^2}\,.
\end{equation}
We remark that in presence of an injected field of amplitude $E_I$ the approximated expression (\ref{Dnew}) also requires $E_I\ll1$.
In Eq. (\ref{Pnew}) the time derivative is always multiplied by $\tau_d$. In order to have the same temporal scaling, when we insert it in Eq. (\ref{E4}) we multiply both sides of that equation by $\tau_d$ and we obtain
\begin{eqnarray}
\tilde{c}\tau_d \frac{\partial E}{\partial z}+\tau_d\frac{\partial E}{\partial t} &=&\frac{\tau_d}{\tau_{p}} \left[-(1+i\theta_c)E+E_{I}+\mu(1-i\alpha)E-(1-i\alpha)|E|^{2}E\right. \nonumber\\
&+&\left. \frac{\tau_{d}^{2}}{(1-i\alpha)} \frac{\partial^{2} E}{\partial t^{2}}
+i\beta |E|^2E-i\tilde{c}\tau_p\frac{k''}{2}\frac{\partial^2 E}{\partial t^2}\right ] \label{E4bis}
\end{eqnarray}
where we have neglected the first order derivative 
$\frac{\tau_d^2}{\tau_p}\frac{\partial E}{\partial t}$ with respect to $\tau_d\frac{\partial E}{\partial t}$ because $\tau_{d}\ll \tau_{p}$.
In the same limit we can replace $\frac{\partial^2 E }{\partial  t^2}$ by $\tilde{c}^2\frac{\partial^2 E }{\partial z^2}$ and thus we have
\begin{eqnarray}
\tilde{c}\frac{\partial E}{\partial z}
+\frac{\partial E}{\partial t} &=&\frac{1}{\tau_p} \left[-(1+i\theta_c)E+E_{I}+\mu(1-i\alpha)E-(1-i\alpha)|E|^2E \right. \nonumber\\
&+&\left.\frac{(\tilde{c} \tau_d)^2}{(1-i\alpha)}
\frac{\partial^2 E}{\partial z^2} 
+i\beta |E|^2E-i \tilde{c}^3\tau_p\frac{k''}{2}\frac{\partial^2 E}{\partial z^2}
\right]\,. \label{E4tris}
\end{eqnarray}
We now pass to a moving reference frame by setting
$z-\tilde{c}t \longrightarrow z $, we define $\theta_0=\theta_c-\mu\beta$, $\Delta=\alpha+\beta$,  $d_R=(\tilde{c}\tau_d)^2/(1+\alpha^2)$ and $d_I=d_R(\alpha+\zeta)$ with $\zeta=-k'' \tilde{c} \tau_{p}\left( 1+\alpha^2 \right)/\left( 2 \tau_d^2 \right)$ and we get
\begin{equation}
\tau_p\frac{\partial E}{\partial t} = E_I-(1+i\theta_0)E+(1-i\Delta)(\mu-|E|^2)E
+(d_{R}+id_I)\frac{\partial^2 E}{\partial z^2} \,. \label{E4new}
\end{equation}
By introducing the scaling
\begin{equation}\label{eq:normal}
\tau =  r t/\tau_{p}\,, \quad
\eta = z \sqrt{ \frac{r}{d_{R}} }\,, \quad
F=  \frac{E} {\sqrt{r}} \,, \quad
F_I = \frac{E_I}{r^{3/2}}\,, \quad
\end{equation}
where $ r= \mu -1 \ll 1$ and setting
\begin{equation}\label{eq:thetabig_zedbar}
\theta = \left( \theta_c+  \mu \alpha \right)/r\,,\quad
G=d_I/d_R\,,
\end{equation}
Eq. (\ref{E4new}) becomes
\begin{eqnarray}\label{eq:LLEactivesupp}
\frac{\partial F}{\partial \tau} &=& F_I + \left(1-i \theta \right)F -  
\left(1-i\Delta\right)|F|^2 F
+ \left (1+i G\right) \frac{\partial^{2} F}{\partial \eta^{2}}\,,
\end{eqnarray}
that coincides with Eq. (3) in the main text with $\gamma=1$. 

Finally, we give an expression of for the detuning parameter, relevant for experiments, i.e. the frequency mismatch of the pump and laser fields.
If we consider Eq. (\ref{eq:LLEactivesupp}) without an injected field ($F_I=0$), the reference frequency $\omega_0$ which appears in the parameter $\theta$ becomes a free parameter which can be conveniently chosen as the free running laser frequency $\omega_L$. This is done by considering the stationary homogeneous solution (obtained by setting $\partial/\partial\tau=\partial^2/\partial\eta^2=0$) of Eq. (\ref{eq:LLEactivesupp}).
By equating separately to zero the real and imaginary parts of the resulting equation we obtain $|F|^2=1$ (in agreement with our scaling) and $\theta=\Delta$ (mode pulling formula). From the previous result and from the definitions of $\theta$ and $\Delta$ it follows 
\begin{equation}\label{eq:omegaL}
    \omega_L=\omega_c+(\alpha-r\beta)\tau_p^{-1}\,,
\end{equation}
which gives the displacement of the laser frequency with respect to the empty cavity frequency. Thus, in presence of an injected field with frequency $\omega_0$, observing that $\theta_c=(\omega_c-\omega_L+\omega_L-\omega_0)\tau_p$, and using the definitions of $\theta$ and $\Delta$ we get:
%
\begin{equation}
    (\omega_L-\omega_0)\tau_p=r(\theta-\Delta)\,.
\end{equation}
Therefore, the pump field is red-detuned (blue-detuned) with respect to the laser field if
$\theta>\Delta$ ($\theta<\Delta$). In this work we study the cavity solitons which require the first condition. We remark, however that in \cite{Prati2020} it was shown that another kind of solitons, the phase solitons, are supported when $\theta<\Delta$.

%
\section{C\lowercase{omparison between the} ESMBE\lowercase{s model and the generalized} LLE \lowercase{approach}}
The model consisting in Eqs. (\ref{E4})-(\ref{P4}) (or ``complete'' model in the following) contains several parameters that in the generalized LLE (\ref{eq:LLEactivesupp}) (or ``reduced'' model in the following) are resumed in the three effective parameters $\theta$, $\Delta$ and $G$. 
In the following we choose the parameters summarized in Table \ref{tab1}, which are compatible with the experiments, that correspond to the effective parameters $\alpha=\Delta=G=2.5$ and $\theta=12$ (and effective cavity length $\eta_\mathrm{max}\sim42$) and compare the results about stability of the homogeneous stationary solutions and spatio-temporal dynamics obtained with the complete and the reduced approaches. 


%
{\renewcommand{\arraystretch}{1.5} 
\begin{table}[h]
\centering
\begin{tabular}{|c|c|c|c|c|c|c|c|c|c|c|c|} \hline
$\theta_c$ & $\alpha$ & $\mu$ & $L$ &  $\tau_d$ & $\tau_p$ & $\tau_e$ & $n$ & $\beta$ & $k''$ \\ 
\hline\hline
-2.4905 & 2.5 & 1.001 & 4.5 mm & 0.1 ps & 100 ps & 0.2 ps & 3.3 & 0 & 0\\
\hline \end{tabular}
\caption{Parametric set for Eqs. (\ref{E4})-(\ref{P4}).}\label{tab1}
\end{table}}
\begin{figure}[t]
    \centering
    \includegraphics[width=0.8\textwidth]{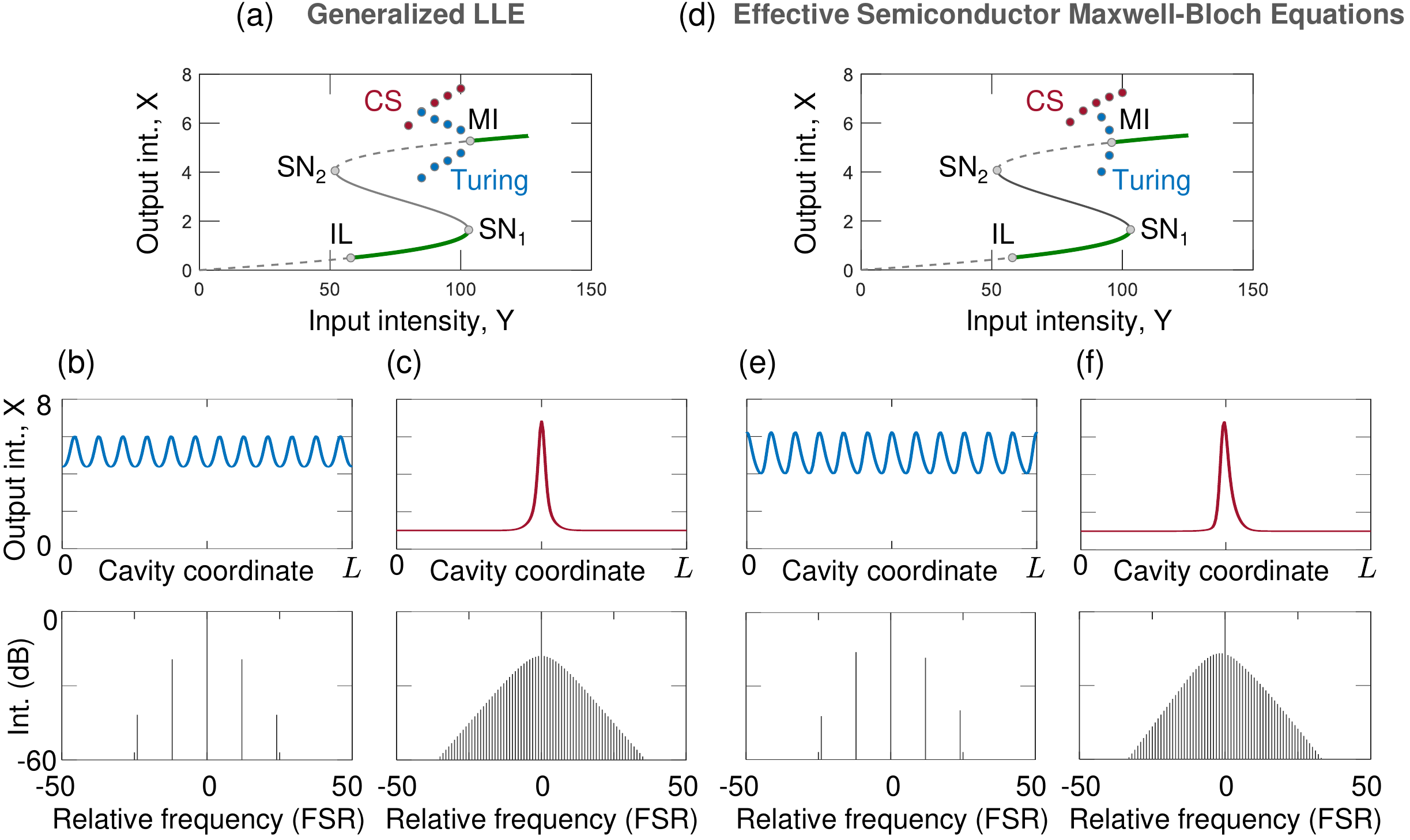}
    \caption{(a)-(c): Results obtained with the reduced model based on Eq. (\ref{eq:LLEactivesupp}) and $\Delta=G=2.5$, and $\theta=12$. (d)-(f): Results obtained with the complete laser model based on Eqs. (\ref{E4})--(\ref{P4}) and the parameters of Table \ref{tab1}.
    In both cases we show the stationary homogeneous solution (solid and dashed lines), the CSs branches (red dots), and the Turing roll branches (blue dots). An example of intensity profile of a Turing pattern (blue curve) for $Y=92$ and a CS for $Y=90$ (red curve), together with the corresponding frequency comb, is also shown for each laser model.}
    \label{fig:comparison}
\end{figure}

Note that for a more direct comparison, we will plot the field intensities in Eqs. (\ref{E4})-(\ref{P4}) with the same scaling adopted in Eq. (\ref{eq:LLEactivesupp}).

Figures \ref{fig:comparison}(a,d) show the two curves of the spatially homogeneous stationary solutions of Eq. (\ref{eq:LLEactivesupp}) (a) and of Eqs. (\ref{E4})--(\ref{P4}) (d) where the output intensity $X=|F|^2$ is plotted versus the input intensity $Y=F_{I}^{2}$ [see Eq. (S13)].
The solid green, dashed grey, and solid grey lines denote stable, unstable, and physically inaccessible states, respectively, and the points IL, SN$_1$, SN$_2$ and MI are 
the bifurcation points defined in the main text. The blue symbols mark the minimum and maximum intensity of the Turing patterns and the red symbols show the maximum intensity for the CS. The stationary curves are indistinguishable. 
The only noticeable difference concerns the modulational instability bifurcation point (MI) whose coordinates are $Y_\mathrm{MI}=95.86$, $X_\mathrm{MI}=5.205$ for Eqs. (\ref{E4})--(\ref{P4}) and $Y_\mathrm{MI}=103.6$, $X_\mathrm{MI}=5.273$ for Eq. (\ref{eq:LLEactivesupp}).

In Figs. \ref{fig:comparison}(b,e) we report an example of Turing pattern for $Y=92$. The upper panel shows the intensity profile along the cavity and the lower panel shows the optical spectrum. The amplitude of the oscillations is larger for the simulation of Eqs. (\ref{E4})--(\ref{P4}) but in both figures the spectrum contains frequencies which are multiples of 12 FSR. Starting from the MI point the pattern bifurcates more rapidly in Fig. \ref{fig:comparison}(d) (the difference between the minima and the maxima values is higher than in the reduced model) and it destabilizes before by decreasing $Y$.

When this occurs we observe the spontaneous formation of a random number of CSs. The CSs can also be switched on by means of an addressing beam as described in Sec. \ref{sec_addressing}. The intensity profile along the cavity and the optical spectrum of a CS for $Y=90$ are shown in Figs. \ref{fig:comparison}(c,f). For the CSs the stable branch is almost identical in the two models. Notice however that the CS obtained with Eqs. (\ref{E4})--(\ref{P4}) is slightly asymmetric, while that of Eq. (\ref{eq:LLEactivesupp}) is perfectly symmetric. We note also that evolutions on the scale of $10^7$ time units in the reduced model confirm the persistence of CSs, both with stable and turbulent background, as stable solutions.

The comparison between Figs. \ref{fig:comparison}(a,d) and Fig. 2(b) of the main text also proves that neglecting the background material nonlinearity and GVD does not affect the overall scenery of the QCL dynamics and the existence of CSs in particular.

At this point we remark that the agreement between the reduced and the complete model in terms of S-shaped curves, linear stability analysis and the spatio-temporal dynamics although very good cannot be perfect. The reason for that is in the evidence that the ESMBEs correctly describe the medium dynamics and thus the associated optical susceptibility both in terms of its frequency dependence and of its nonlinear dependence on the intracavity field intensity, as opposed to the reduced model derived in the hypothesis of ``fast'' medium and vicinity to the lasing threshold operation (cubic nonlinearity). 
While this implies different positions of the MI, SN and IL points and a slightly modified bifurcation scenery of the spatially modulated solutions, we can anyway confirm that the generalized LLE model is excellently predictive of the laser dynamics, compared to the complete ESMBEs model; in particular it is able to reproduce the existence of stable Turing rolls and localized solutions in the form of CSs under the same conditions.

\begin{figure}[t!]
    \centering
    \includegraphics[width=0.5\textwidth]{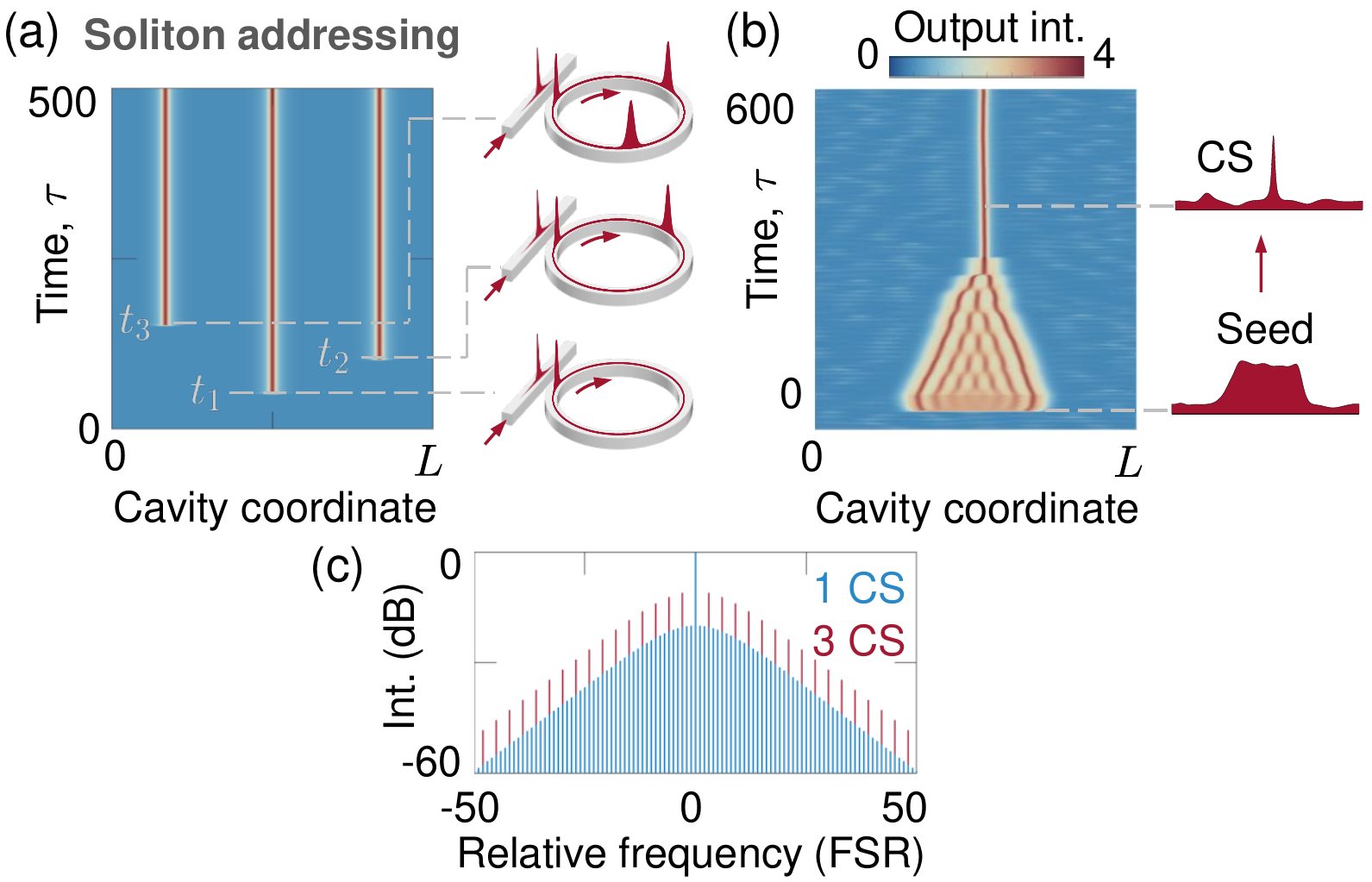}
    \caption{\textbf{Spectral shaping by external writing of temporal solitons.} (a) Spatio-temporal plot showing the evolution of the intracavity pattern in a ring QCL as multiple cavity solitons (CSs) are sequentially excited by the injection of pulses at times $t_1$, $t_2$ and $t_3$. (b) Excitation of a single CS with a wide pulse seed, approximately 20 times wider than the CS. (c) Frequency comb spectra taken at the times $t_1$ and $t_3$ in (a), corresponding to the excitation of a single CS and three equidistant CSs, respectively.}
    \label{fig_addressing}
\end{figure}

\section{S\lowercase{oliton addressing}}
\label{sec_addressing}

In Fig. \ref{fig_addressing}(a) we show a sequence of three pulses (width comparable to CS) being subsequently injected through the external addressing channel (here schematized  by a straight waveguide evanescently coupled to the ring QCL) and turning on three CSs in different positions of the resonator. The parameter set used for these simulations is the same one used to produce the results of Fig. 2 of the main text.
The system is initially in the lower branch uniform solution and the first short pulse excites a CS at $z=L/2$. After 700 roundtrips a second identical pulse excites a second CS at $z=5L/6$ and after another 700 roundtrips the same is repeated to create a third CS at $z=L/6$. The CSs are independent and unperturbed one from another. Figure \ref{fig_addressing}(b) shows how a much broader pulse can be used to create a single CS, which may be convenient in practical realizations. The optical spectra corresponding to the initial and final configurations demonstrates how the addition of CSs allows to manipulate the OFC (Fig. \ref{fig_addressing}(c)). In particular, exciting equally-spaced CSs allows to control the spacing between the OFC lines and realize an externally addressable harmonic frequency comb \cite{Kazakov2017}. Besides this, non-equidistant CSs may also be excited, giving a powerful degree of freedom in shaping OFCs. Further simulations confirm that the CSs remain independent within a spatial separation comparable to the period of the global structure bifurcated at the MI point \cite{Brambilla1996}, i.e. similar to the pitch of the rolls of Fig. 2(d) of the main text. Moreover we verified that the CSs can drift unperturbed along gradients in the input field $F_I$, hence they are spatially reconfigurable.

\section{L\lowercase{ocalized structures in the} G\lowercase{eneralized} L\lowercase{ongitudinal} LLE: A \lowercase{unifying picture}}
\begin{figure}
    \centering
    \includegraphics[width=0.5\textwidth]{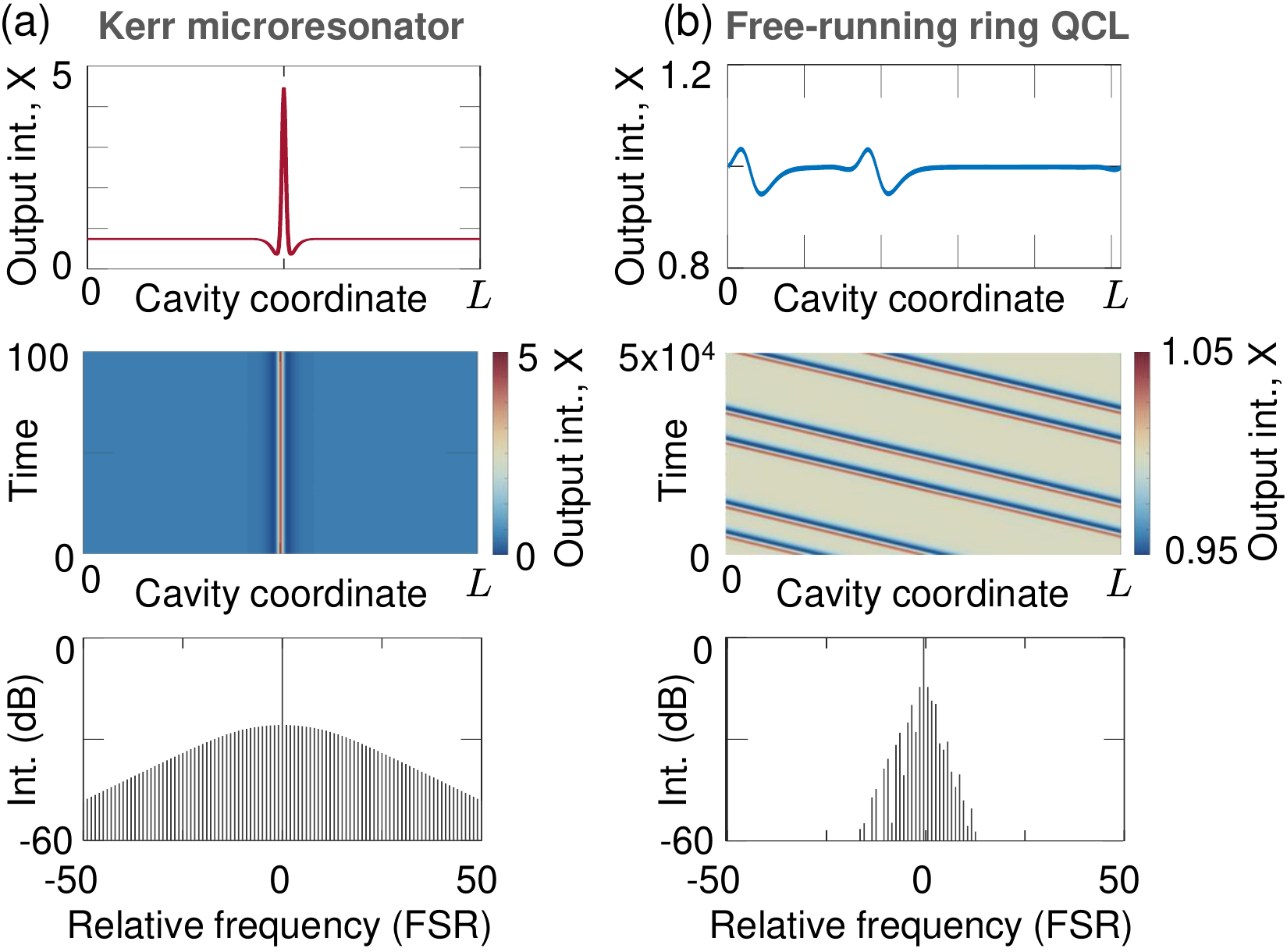}
    \caption{Intracavity intensity, spatiotemporal plots and corresponding frequency spectra of: (a) a Kerr soliton from a microresonator; (b) the localized structures from a free-running ring QCL similar to those reported in Extended Data Fig. 1b of Ref. \citenum{Piccardo2020}. Time is scaled on $\tau_{p}$.}
    \label{fig:s3}
\end{figure}

We named Eq. (1) in the main text \lq\lq Generalized Longitudinal LLE" because 
under particular assumptions it reduces to the longitudinal LLE for a Kerr resonator but with other choices of the parameters it can as well describe the pattern formation and structures localization occurring in the other two systems shown in Fig. 1.\\
%
%
Equation (1) reduces to the LLE in the case of a passive medium where the dispersive effects are dominant and absorption is small. For instance one can set in Eq. (1) $\Delta=-100$, $d_I=100d_R$, and $\mu=-0.01$, with the additional choice $\theta_0=1$ (see Table S2).
Since $\mu\Delta=1$, this amounts to $\theta=\theta_0+\mu\Delta=2$ in Eq. (2) and it is well known that in the LLE the output is bistable and CSs exist if $\theta>\sqrt{3}$. An example of CS for $E_I=1.385$ and cavity length $L=100\sqrt{d_R}$ is shown in Fig. \ref{fig:s3}(a). 
Notice that this result was obtained by integrating Eq. (1) without making the approximations $\mu(1-i\Delta)\approx-i\mu\Delta$ and $d_R+id_I\approx id_I$ introduced in the main text to pass from Eq. (1) to the LLE, i.e. Eq. (2).

In the other two examples shown in Fig. 1 of the main text the localized structures arise in a ring QCL kept few percent above threshold, with or without a driving field.
The new case of a driven QCL was the main object of the present paper. The results presented for instance in Fig. 2 of the main text were
obtained by integrating Eq. (3) but, taking into account the various changes of variables, they can be equivalently obtained by integrating Eq. (1) provided one specifies how much above threshold the laser is. For instance, 
for a laser $1\%$ above threshold, i.e. $\mu=1.01$, the other parameters are listed in the second row of Tab. \ref{tab:s2}, where in particular the value of $E_I$ corresponds to the value $Y_4$ of Fig. 2 and to the soliton shown in Fig. 2(c).  Notice that the frequency of the driving field in this case is very close to that of the solitary laser because there would be perfect matching if $\Delta+\theta_0=0$
(the relation $\Delta+\theta_{0}=0$ is equivalent to the relation $\Delta=\theta$ which ensures that the frequency  $\omega_{L}$ of the free running laser coincides with the frequency $\omega_{0}$ of the injected field) and here $\Delta+\theta_0=0.027$. This circumstance allows to find stable CSs even for a very small amplitude of the injected field. Another condition \cite{Prati2020} that favors the existence of CSs is $G\equiv d_I/d_R > \Delta$. In a QCL with negligible Kerr effect $\Delta=\alpha$ and $G=\alpha+\zeta$, where $\alpha$ is the linewidth enhancement factor and $\zeta$ is a dimensionless parameter related to GVD, therefore $\zeta$ must be positive ($k''$ negative).

\begin{figure}
    \centering
    \includegraphics[width=0.75\textwidth]{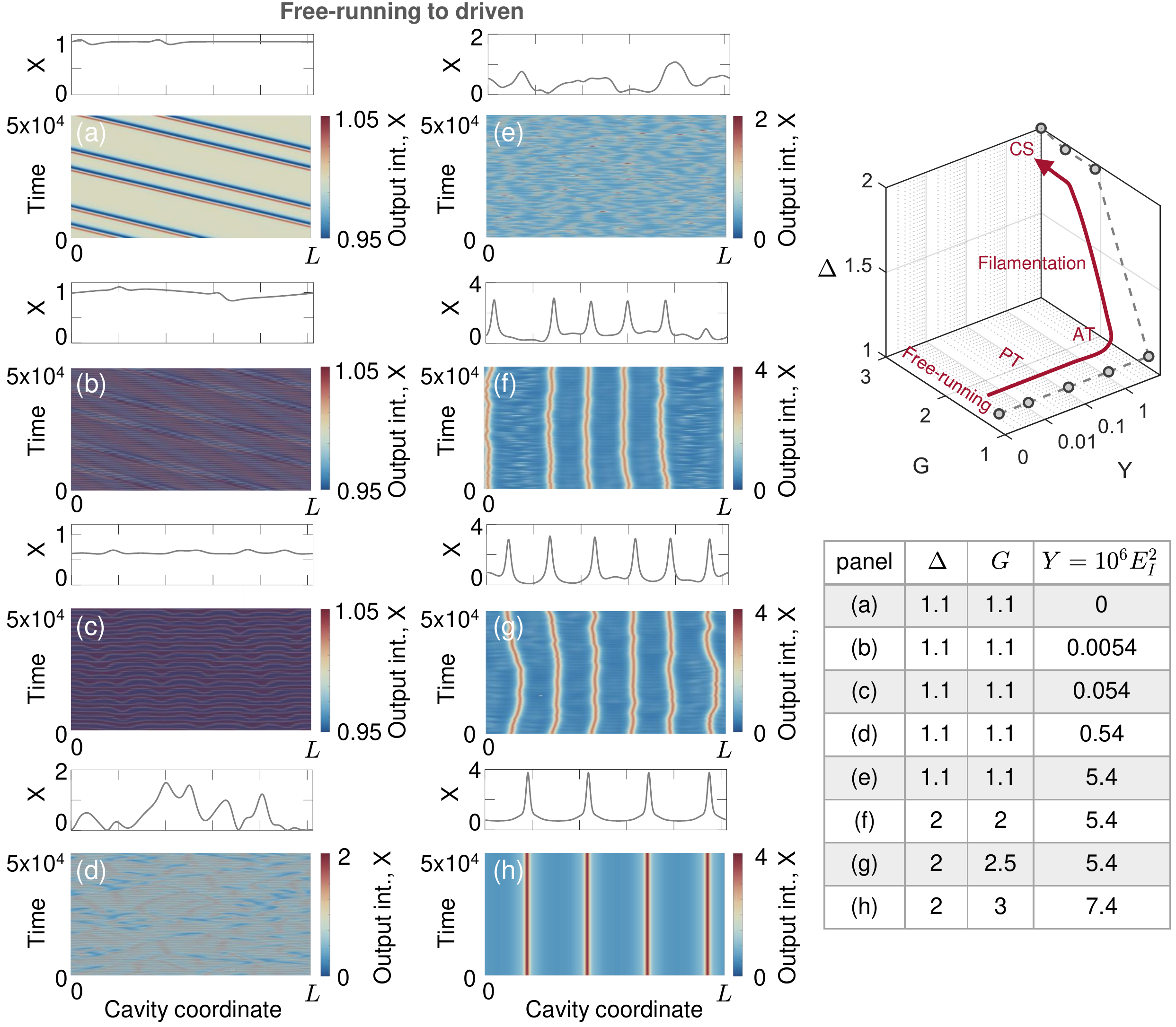}
    \caption{(a)--(h), Spatiotemporal plots of different dynamical regimes obtained when sweeping the multidimensional parameter space along the trajectory shown on the right plot (the corresponding parameter values are reported in the table). The starting point is represented by the localized structures of the free-running case ($Y=0$) shown in Fig. \ref{fig:s3}b,
 and the final point corresponds to a train of CSs identical to the one reported in Fig. 2(c). PT, phase turbulence; AT, amplitude turbulence. (a), (b), (e), (g), (h) correspond to the cases of the 1D profiles shown in the box of Fig. 2 of the main text (top to bottom, respectively). Time is scaled on $\tau_{p}$, while the other parameters are: $\mu=1.01$, $L/\sqrt{d_{R}}=1000$, $\theta=(\theta_{0}+\mu\Delta)/(\mu-1)= 4.7$.}
    \label{fig:ssmooth}
\end{figure}

{\renewcommand{\arraystretch}{1.5} 
\begin{table}[]
    \centering
    \begin{tabular}{|c||c|c|c|c|c|c|c|}
    \hline
                    & $E_I$ & $\mu$ & $\theta_0$ & $\Delta$ & $G=d_I/d_R$ & $L/\sqrt{d_R}$ \\
    \hline\hline
    Kerr microresonator               & 1.385     &   -0.01        & 1 &   -100   & 100 &  100 \\
    \hline
    Injected ring QCL          & $\sqrt{7.4}\times10^{-3}$ &   1.01 & -1.973 & 2 & 3 &  1000 \\
    \hline
    Free-running ring QCL    & 0     &   1.01        & -1.064 &   1.1   & 1.1 &  1000 \\
    \hline
    \end{tabular}
    \caption{Example of three sets of parameters corresponding to the three systems sketched in Fig. 1 where localized structures were found by numerical integration of the Generalized Longitudinal LLE.}
    \label{tab:s2}
\end{table}}

In absence of the driving field the output can never be bistable and CSs of the type considered in the two previous systems, which can be regarded as portions of a modulated pattern on a homogeneous background, cannot exist anymore. In this case Eq. (3) reduces to a CGLE which predicts the existence of different types of localized structures emerging from the Benjamin-Feir instability for $G\Delta>1$. In particular, when $G\Delta$ is larger but close to $1$ the instability may produce localized structures in the form of small amplitude fluctuations superimposed to the homogeneous state and associated with large phase fluctuations (phase instability) \cite{Aranson2002}. These coincide with the localized structures which exhibit a $\sech^2$ spectrum recently reported in Ref. \citenum{Piccardo2020}. As shown in Fig. \ref{fig:s3}(b), similar structures were obtained directly from Eq. (1) with the parameters reported in Tab. \ref{tab:s2}. \\
Moreover, we show that it is possible to pass from the localized shallow structures of the free-running case reported in Fig. \ref{fig:s3}(b) to the high contrast CSs of the driven case reported in Fig. 2(c) by sweeping an appropriate trajectory in the multidimensional parameter space (Fig. \ref{fig:ssmooth}). In fact, with reference to Fig. \ref{fig:ssmooth}, starting with localized solutions corresponding to $Y=0$ (a) and increasing the value of $Y$ (while keeping the other parameters specified in the caption unchanged), we observe the system to evolve first towards an irregular regime characterized by small amplitude fluctuations around a constant intensity values (phase instability, panels b, c) and then on an irregular regime characterized by large amplitude fluctuations (amplitude instability, panels d,e). At this stage, if the values of $\Delta$ and $G$ are increased the output becomes bistable and the amplitude turbulence seeds spontaneous formation of CSs on a turbulent (panels f,g) and stable (panel h) background. In particular,
Figure \ref{fig:ssmooth}(h) shows an array of CSs identical to the one in Fig. 2(c).

These results clearly show the link among the different types of temporal localized structures/frequency combs reported so far in both passive Kerr microresonator and ring QCLs that all appear as self-organized solutions of the ``Generalized Longitudinal LLE'' in different regions of its parameter space.